\documentclass[aps,prb,showpacs,twocolumn,citeautoscript,superscriptaddress]{revtex4}
\usepackage{amsmath}
\usepackage{graphicx}
\usepackage{dcolumn}
\usepackage{float}

\begin{document}
\title{Magnetic and transport properties of Pr$_{2}$Pt$_{3}$Si$_{5}$}
\author{V. K. Anand}
\email{vanand@ameslab.gov; Present address: Ames Laboratory, Department of Physics and Astronomy, Iowa State University, Ames, Iowa 50011, USA.}
\affiliation {Department of Condensed Matter Physics and Materials Science, Tata Institute of Fundamental Research, Mumbai-400005, India}

\author{Anupam}
\affiliation{Department of Physics, Indian Institute of Technology, Kanpur-208016, India}

\author{Z. Hossain}
\affiliation{Department of Physics, Indian Institute of Technology, Kanpur-208016, India}

\author{S. Ramakrishnan}
\affiliation {Department of Condensed Matter Physics and Materials Science, Tata Institute of Fundamental Research, Mumbai-400005, India}

\author{A. Thamizhavel}
\affiliation {Department of Condensed Matter Physics and Materials Science, Tata Institute of Fundamental Research, Mumbai-400005, India}

\author{D. T. Adroja}
\affiliation{ISIS Facility, Rutherford Appleton Laboratory, Chilton, Didcot Oxon, OX11 0QX, UK}

\date{\today}

\begin{abstract}

We have investigated the magnetic and transport properties of a polycrystalline Pr$_{2}$Pt$_{3}$Si$_{5}$ sample through the dc and ac magnetic susceptibilities, electrical resistivity, and specific heat measurements. The Rietveld refinement of the powder X-ray diffraction data reveals that Pr$_{2}$Pt$_{3}$Si$_{5}$ crystallizes in the U$_{2}$Co$_{3}$Si$_{5}$-type orthorhombic structure (space group \textit{Ibam}). Both the dc and ac magnetic susceptibility data measured at low fields exhibit sharp anomaly near 15 K. In contrast, the specific heat data exhibit only a broad anomaly implying no long range magnetic order down to 2 K. The broad Schottky-type anomaly in low temperature specific heat data is interpreted in terms of crystal electric field (CEF) effect, and a CEF-split singlet ground state is inferred. The absence of the long range order is attributed to the presence of nonmagnetic singlet ground state of the Pr$^{3+}$ ion. The electrical resistivity data exhibit metallic behavior and are well described by the Bloch-Gr\"{u}niesen-Mott relation.

\end{abstract}


\maketitle

\section{Introduction}

The rare earth ternary intermetallic compounds R$_{2}$T$_{3}$X$_{5}$ (R = rare-earth, T = transition metal, and X = Si, Sn or Ge) are well known for their diverse physical properties such as Kondo lattice behavior, heavy-fermion behavior, valence fluctuation, unusual magnetic order and superconductivity [1-22]. For example, among the Ce-based 235 compounds, Ce$_{2}$Ni$_{3}$Ge$_{5}$ is an antiferromagnetically ordered Kondo lattice system that exhibits pressure induced superconductivity around 3.9 GPa \cite{18,19}, Ce$_{2}$Ni$_{3}$Si$_{5}$ and Ce$_{2}$Co$_{3}$Ge$_{5}$ exhibit valence fluctuation \cite{20,21}, and Kondo lattice compounds Ce$_{2}$Rh$_{3}$Ge$_{5}$ and Ce$_{2}$Ir$_{3}$Ge$_{5}$ exhibit moderate heavy fermion behavior in their antiferromagnetic ground state \cite{22}. In contrast to the Ce-compounds, where electronic ground state is determined by the competition between RKKY and Kondo interactions, in Pr-compounds, the ground state critically depends on the crystal electric field (CEF) effect. The possibility that the nine-fold degenerate ground states of Pr (J = 4) split into a combination of CEF-split states with a singlet state makes the study of Pr-compounds very interesting. An interesting consequence of the CEF-split singlet state in Pr-compounds is the realization of heavy-fermion behavior, as in PrOs$_{4}$Sb$_{12}$ \cite{23,24} and PrRh$_{2}$B$_{2}$C \cite{25}, and spin-glass behavior in PrAu$_{2}$Si$_{2}$ \cite{26,27} and PrRuSi$_{3}$ \cite{28}. The spin-glass behavior in singlet ground state Pr-compounds is proposed to originate from the dynamic fluctuations of low lying crystal field levels that frustrates the induced moment magnetism in these compounds \cite{27}.

Continuing our work on Pr-based system Pr$_{2}$T$_{3}$X$_{5}$ (T = transition metal, X = Si, Ge), we have investigated a new compound of this series, Pr$_{2}$Pt$_{3}$Si$_{5}$. Recently, we have reported the magnetic and transport properties of Pr$_{2}$T$_{3}$Ge$_{5}$ (T = Ni, Co, Pd, Rh) and Pr$_{2}$Pd$_{3}$Si$_{5}$ \cite{21,29,30,31}. We found that both Pr$_{2}$Ni$_{3}$Ge$_{5}$ and Pr$_{2}$Pd$_{3}$Ge$_{5}$ order antiferromagnetically below 8.5 and 8.3 K, respectively, and exhibit anomalously high positive magnetoresistance ($\sim$ 900\% in Pr$_{2}$Pd$_{3}$Ge$_{5}$ at 2 K and 10 T) \cite{29,30}. On the other hand, Pr$_{2}$Rh$_{3}$Ge$_{5}$ and Pr$_{2}$Co$_{3}$Ge$_{5}$ remain paramagnetic down to 0.5 K and 2 K respectively, with a singlet ground state \cite{21,30}. The excitonic mass enhancement of the effective mass due to low lying crystal field excitations is believed to be responsible for a moderate heavy fermion state in Pr$_{2}$Rh$_{3}$Ge$_{5}$. Here, we report the synthesis and investigation of magnetic and transport properties of a new ternary intermetallic compound Pr$_{2}$Pt$_{3}$Si$_{5}$. The Ce-analog, Ce$_{2}$Pt$_{3}$Si$_{5}$ has already been characterized as a Kondo lattice system that undergoes an antiferromagnetic transition below 6.3 K \cite{32,33}.

\section{Experimental}

The polycrystalline sample of Pr$_{2}$Pt$_{3}$Si$_{5}$ was prepared by standard arc melting of high purity elements (Pr-99.9\%, Pt-99.99\% and Si-99.999\%) on a water cooled copper hearth under an inert argon atmosphere. In order to achieve better homogeneity, the sample was flipped and re-melted several times during the arc melting process. Further, to improve the phase formation, the arc melted button was sealed inside the quartz tube under vacuum and annealed at 1000$^{o}$C for a week. The crystal structure and phase purity of the sample were determined by X-ray diffraction and high resolution SEM images. The dc magnetic susceptibility was measured by using a commercial SQUID magnetometer (MPMS, Quantum-design, USA). The specific heat was measured by relaxation method in a physical properties measurement system (PPMS, Quantum-design, USA). The electrical resistivity was measured by conventional four probe technique using an LR700 resistance bridge (Linear Research, USA). The ac susceptibility measurements were also performed using the PPMS (Quantum-design, USA).

\begin{figure}
\centering
\includegraphics[width=8cm, keepaspectratio]{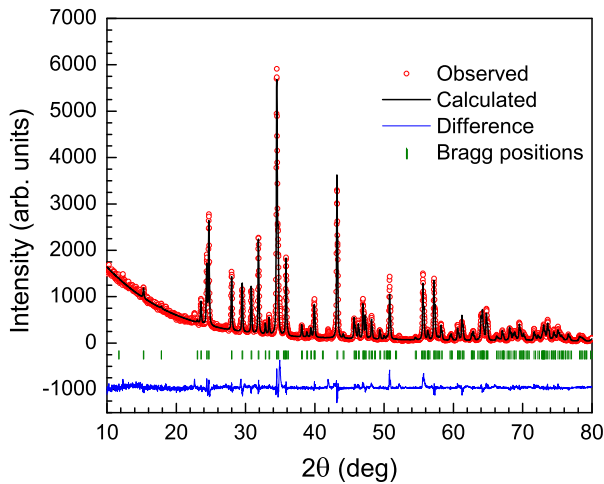}
\caption{\label{fig1} (Colour online) Powder X-ray diffraction pattern of Pr$_{2}$Pt$_{3}$Si$_{5}$ recorded at room temperature. The solid line through the experimental points is the Rietveld refinement fit calculated with U$_{2}$Co$_{3}$Si$_{5}$-type orthorhombic (space group \textit{Ibam}) structural model. The vertical short bars indicate the Bragg positions. The lower most curve represents the difference between the experimental and model results.}
\end{figure}

\begin{table} 
\caption{\label{tab:table1} The crystallographic parameters of Pr$_{2}$Pt$_{3}$Si$_{5}$ obtained from the Rietveld structural refinement of powder X-ray diffraction data using Fullprof.  The least squares refinement quality parameter $\chi^{2}$ = 4.2.}
\begin{tabular}{llll}
\hline

Structure & \multicolumn{3}{l}{U$_{2}$Co$_{3}$Si$_{5}$-type orthorhombic}  \\
Space group & \textit{Ibam} (No. 72) \\
Lattice  parameters \\
 \hspace{1cm} a & 9.9434 (3) \AA \\
 \hspace{1cm} b & 11.5961(4) \AA \\
 \hspace{1cm} c & 6.0495(2) \AA \\
 \hspace{1cm} V$_{cell}$  & 697.53(4) \AA $^{3}$ \\
Atomic Coordinates \\
 \hspace{0.8cm}  Atom (site) &	 x 		    &	y			&	z    \\			
  \hspace{1cm}     Pr (8j)   &	0.2651(3) 	&	0.1283(3)	&   0	 \\
  \hspace{1cm}     Pt1 (4a)	 &	0 			&	0			&	0.25 \\
  \hspace{1cm}     Pt2 (8j)	 &	0.1116(2) 	&   0.3639(2) 	&	0    \\
  \hspace{1cm}     Si1 (4b)	 &	0.5	 		&	0			&	0.25 \\
   \hspace{1cm}    Si2 (8g)	 &	0			&	0.2239(13) 	&	0.25 \\
   \hspace{1cm}    Si3 (8j)	 &	0.3511(13) 	& 	0.3791(15)	& 	0	 \\
\hline
\end{tabular}

\end{table}

\section{Results and Dscussion}

The X-ray diffraction (XRD) data collected on the powdered Pr$_{2}$Pt$_{3}$Si$_{5}$ sample were analyzed by the Rietveld structural refinement using a freely available software (Fullprof \cite{34}). The powder XRD pattern of Pr$_{2}$Pt$_{3}$Si$_{5}$ together with the Rietveld refinement profile fit is shown in Fig.~1. An ordered crystallographic structure and single phase nature of sample are evident from the refinement of the powder XRD data. From the Rietveld refinement, we found that Pr$_{2}$Pt$_{3}$Si$_{5}$ crystallizes in the U$_{2}$Co$_{3}$Si$_{5}$-type orthorhombic (space group \textit{Ibam}). The crystallographic parameters obtained from the best fit of Rietveld refinement using least squares method are tabulated in Table~1. The lattice parameters a = 9.943~Å, b = 11.596~Å and c = 6.049~Å of Pr$_{2}$Pt$_{3}$Si$_{5}$ are smaller than and comparable to that of La$_{2}$Pt$_{3}$Si$_{5}$ and Ce$_{2}$Pt$_{3}$Si$_{5}$ \cite{32,33}. The unit cell volumes of La$_{2}$Pt$_{3}$Si$_{5}$, Ce$_{2}$Pt$_{3}$Si$_{5}$ and Pr$_{2}$Pt$_{3}$Si$_{5}$ are 727.04, 725.9 and 697.53~Å$^{3}$, respectively. Few weak un-indexed peaks are also present in XRD pattern which could not be identified, and the unidentified impurity phase(s) is estimated to be less than 4\%. The phase purity was further checked through the high resolution SEM images which also confirmed the almost single phase nature of the sample.

\begin{figure}
\centering
\includegraphics[width=8cm, keepaspectratio]{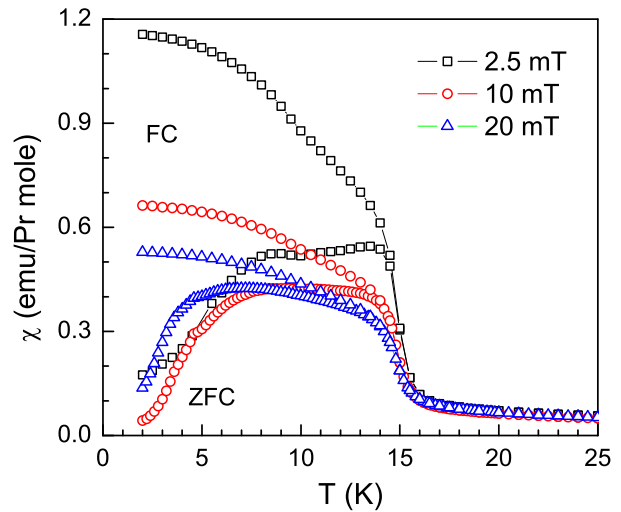}
\caption{\label{fig2} (Colour online) The low temperature zero field cooled (ZFC) and field cooled (FC) dc magnetic susceptibility $\chi(T)$ of Pr$_{2}$Pt$_{3}$Si$_{5}$ as a function of temperature in the temperature range 2 -- 25 K measured in low magnetic fields of 2.5, 10 and 20 mT.}
\end{figure}

\begin{figure}
\centering
\includegraphics[width=8cm, keepaspectratio]{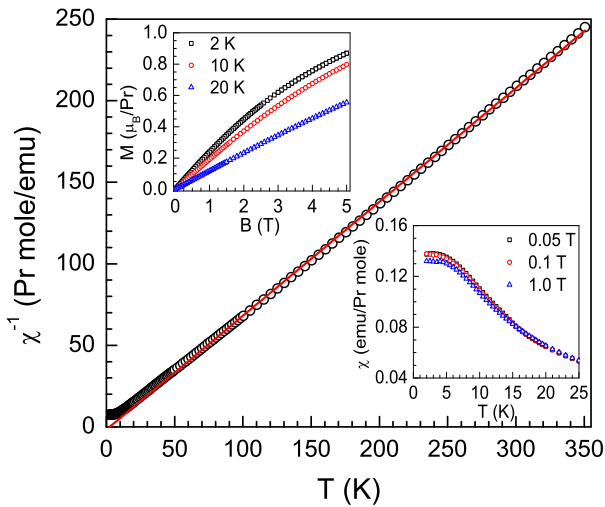}
\caption{\label{fig3} (Colour online) The inverse dc magnetic susceptibility, $\chi^{-1}(T)$ of Pr$_{2}$Pt$_{3}$Si$_{5}$ as a function of temperature in the temperature range 2 -- 350 K measured in a field of 1.0 T. The solid line represents the fit to Curie-Weiss law. The lower inset shows the low temperature ZFC susceptibility measured at different applied magnetic fields. The upper inset shows the dc isothermal magnetization $M(B)$ as a function of the magnetic field measured at  2, 10 and 20  K.}
\end{figure}

The dc magnetic susceptibility and magnetization data of Pr$_{2}$Pt$_{3}$Si$_{5 }$ measured as a function of temperature and magnetic field are presented in Fig.~2 and Fig.~3. As shown in Fig.~2, at low fields (e.g., at 2.5~mT), zero field cooled (ZFC) magnetic susceptibility data exhibit a sharp increase below 16~K with a broad peak over a temperature range of  15 to 8~K. A hysteresis is also observed between the zero field cooled (ZFC) and field cooled (FC) susceptibility data with a split up at 15~K. An increase in the magnetic field strength leads to a reduction in the magnitude of susceptibility and eventually suppresses the 15 K susceptibility anomaly. At a field of 0.05~T (and higher), susceptibility tends to attain a constant value at low temperatures (lower inset of Fig.~3).  At high temperatures, a Curie-Weiss behavior, $\chi = C/(T-\theta_{p})$ is observed in the susceptibility data. From a linear fit to the inverse magnetic susceptibility data (measured at 1.0~T) above 100~K (the solid line in Fig.~3) we obtain an effective moment of $\mu_{eff}$ = 3.39~$\mu_{B}$ and Curie-Weiss temperature, $\theta_{p \-}$ = +2.4~K. The effective moment obtained is close to the theoretical value of the effective moment of Pr$^{3+}$ ion (3.58 $\mu_{B}$).

The magnetic field dependence of the isothermal magnetization is shown in the upper inset of Fig.~3. At 2 and 10~K, which are well below the temperature of the observed susceptibility anomaly, the isothermal magnetization exhibits a small nonlinearity which could be due to the presence of the short range correlations and/or crystal electric field effects. The nonlinear behavior of isothermal magnetization at 2 and 10~K may also arise from the spin glass type behavior as evidenced from the thermo-remnant magnetization discussed below. At 20~K, which is above the temperature of the susceptibility anomaly, a linear field dependence is observed in the isothermal magnetization. The magnetization does not reach its theoretical value of M$_{s}$ = 3.2~$\mu_{B}$ for Pr$^{3+}$ ions up to the maximum measured field of 5~T.

\begin{figure}
\centering
\includegraphics[width=8cm, keepaspectratio]{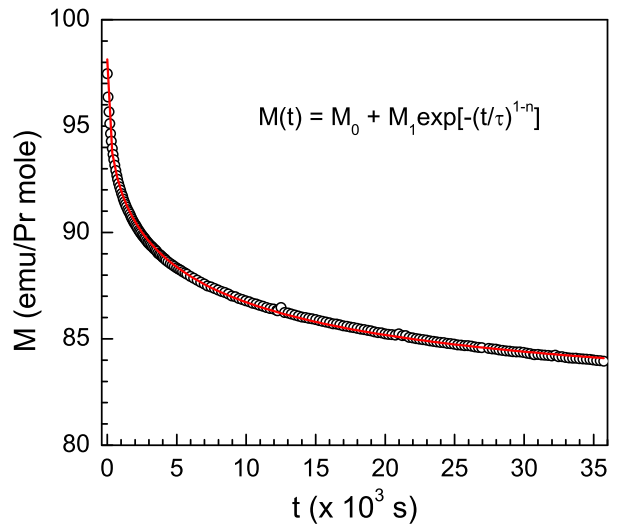}
\caption{\label{fig4} (Colour online) The time dependence of thermo-remnant magnetization (TRM), $M(t)$ of Pr$_{2}$Pt$_{3}$Si$_{5}$ recorded at 2 K after switching off the cooling magnetic field of 0.05 T. The solid line is the fit to the superposition of a stretched exponential and a constant term (see text).}
\end{figure}

We also measured the time dependence of thermo-remnant magnetization (TRM), $M(t)$, at 2 K and it is shown in Fig.~4. The sample was cooled in a magnetic field of 0.05~T from 50 K to 2~K and the field-cooled isothermal remnant magnetization was measured after switching off the magnetic field. A superposition of a stretched exponential and a constant term, $M = M_{0} + M_{1} exp[-(t/\tau)^{1-n}]$ was used to fit the observed TRM data (solid line in Fig.~4). The best fit was obtained with parameters $M_{0}$ = 82.13 emu/mole, $M_{1}$ = 16.01 emu/mole, mean relaxation time $\tau$ = 5788 s and n = 0.59. The values of $\tau$ and $n$ are typical for a spin glass system. The constant term $M_{0}$ is interpreted as the longitudinal spontaneous magnetization coexisting with the frozen transverse spin component \cite{35,36}.

\begin{figure} 
\centering
\includegraphics[width=8cm, keepaspectratio]{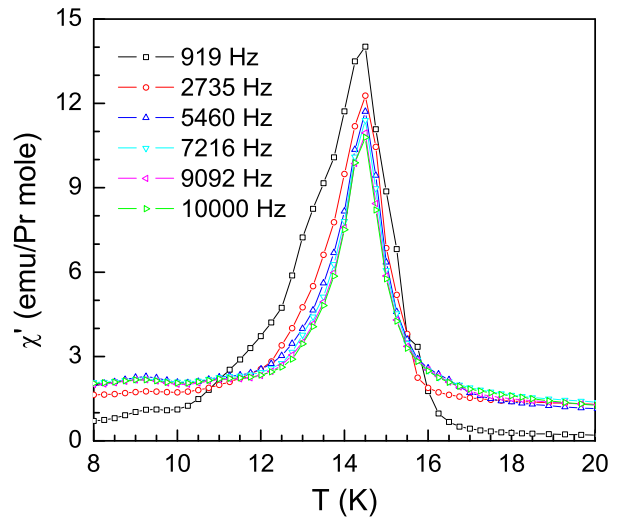}
\caption{\label{fig5} (Colour online) The temperature dependence of the real part of ac magnetic susceptibility of Pr$_{2}$Pt$_{3}$Si$_{5}$ measured at different frequencies.}
\end{figure}

\begin{figure}
\centering
\includegraphics[width=8cm, keepaspectratio]{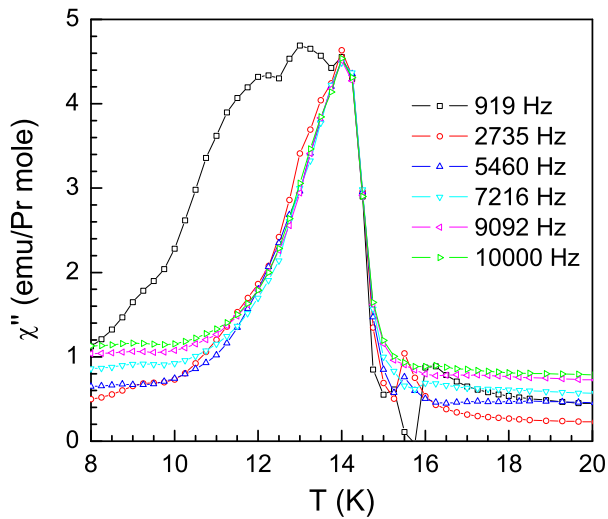}
\caption{\label{fig6} (Colour online) The temperature dependence of the imaginary part of ac magnetic susceptibility of Pr$_{2}$Pt$_{3}$Si$_{5}$ measured at different frequencies.}
\end{figure}

The observations of both splitting of ZFC and FC magnetic susceptibility and slow relaxation of magnetization are consistent with a spin-glass behavior; however, a ferromagnetic impurity can also produce such a behavior. We are not aware of any binary or ternary compound made from Pr-Pt-Si combination which is reported to exhibit an anomaly near 15 K. Therefore, in order to understand the nature of magnetic susceptibility anomaly in Pr$_{2}$Pt$_{3}$Si$_{5}$, we have also measured the ac magnetic susceptibility at different frequencies. The real and imaginary parts of ac magnetic susceptibility measured in an excitation field of 1.5~mT are shown in Fig.~5 and Fig.~6, respectively. Both real and imaginary parts of the ac magnetic susceptibility exhibit a sharp peak near 15 K, however, no frequency dependence is observed in the ac susceptibility anomaly which is contrary to the behavior expected for a canonical spin-glass system. It is observed that at lower frequencies (e.g., at 919 Hz) the ac susceptibility peak in the imaginary part becomes broader whereas, that in the real part increases in magnitude.

\begin{figure}
\centering
\includegraphics[width=8cm, keepaspectratio]{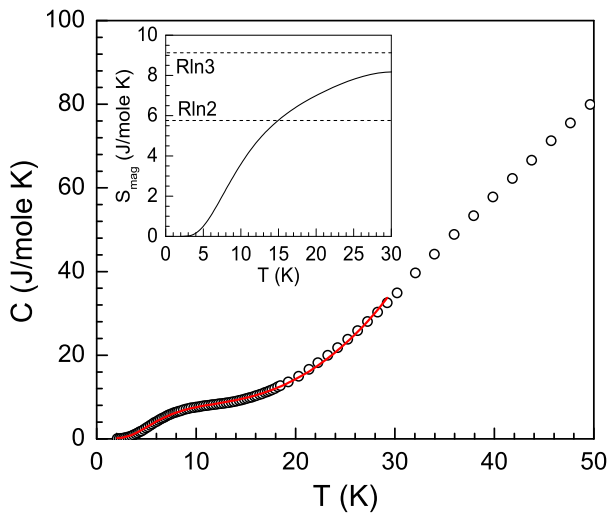}
\caption{\label{fig7} (Colour online) The temperature dependence of specific heat data of Pr$_{2}$Pt$_{3}$Si$_{5}$ measured in different magnetic fields. The solid line represents the fit to $C = \gamma T + \beta T^{3} + C_{Sch}$ described in text. The inset shows the magnetic contribution to entropy.}
\end{figure}

The specific heat data of Pr$_{2}$Pt$_{3}$Si$_{5}$ measured in zero magnetic field are shown in Fig.~7. The specific heat data exhibit only a broad hump near 9 K, no sharp anomaly is observed corresponding to the 15~K anomaly in the magnetic susceptibility. The broad Schottky-type anomaly observed in the specific heat can be attributed to the crystal field effect (CEF). At 2~K, the specific heat has a very small value of 46.4~mJ/mole~K, which suggests a CEF-split singlet electronic ground state in Pr$_{2}$Pt$_{3}$Si$_{5}$. The specific heat data below 30~K were analyzed using $C = \gamma T + \beta T^{3} + C_{Sch}$, a fit to this expression is shown by a solid line in Fig.~7. The best fit was obtained for $\gamma$ = 56~mJ/mole\,K$^{2}$ and $\beta$ = 1.21~mJ/mole\,K$^{4}$.

The crystal field contribution $C_{Sch}$ involves three crystal field levels. The ground state is a singlet, first and second excited states are also singlets situated at 22~K and 35~K above the ground state.  The temperature dependence of the magnetic contribution to entropy obtained from the analysis of specific heat is shown in the inset of Fig.~7. The magnetic entropy attains a value of $R \ln2$ near 15~K which reveals a singlet ground state. The CEF-split singlet ground state 235 compound ${\rm Pr_2Co_3Ge_5}$ is found to have a singlet first excited state at 22~K and second excited doublet at 73~K from the crystal field analysis of the specific heat data \cite{21}. ${\rm Pr_2Rh_3Ge_5}$ is another CEF singlet ground state compound for which the specific heat data indicate four singlets at 0, 12, 40 and 60~K \cite{30}. The inelastic scattering of conduction electrons by low lying crystal field levels in ${\rm Pr_2Rh_3Ge_5}$ is believed to lead to a moderate heavy fermion state. The CEF-split singlet ground state in the present compound ${\rm Pr_2Pt_3Si_5}$ can be held responsible for the absence of long range magnetic order. A singlet ground state system can exhibit a self-induced spontaneous moment ordering only if the ratio $J_{ex}/\Delta$ is above a critical value, where  $J_{ex}$ is the exchange integral and $\Delta$ is the splitting energy between the ground state and the first excited state. Therefore further investigations using neutron scattering and muon spin relaxation ($\mu$SR) are called for to explore whether the observed magnetic susceptibility anomaly is due to the induced moment magnetism behavior. This will help in understanding the possibility of spin-glass behavior by the dynamic fluctuations of crystal field level in ${\rm Pr_2Pt_3Si_5}$.

The Debye temperature estimated using the relation $\Theta_{D} = (12\pi^{4} N_{A} r K_{B}/5\beta)^{1/3}$ (where $r$ is the number of atoms per formula unit) came out to be $\Theta_{D}$ = 252~K. The value of Debye temperature $\Theta_{D}$ = 252~K obtained is comparable to that of $\Theta_{D}$ = 273~K for Pr$_{2}$Ni$_{3}$Ge$_{5}$ ($\beta$ = 0.96~mJ/mole\,K$^{4}$). In Pr-compounds because of the presence of crystal field effect the precise estimate of specific heat coefficient $\beta$ is little difficult. However, the specific heat of 235 La-compounds can be used to estimate the coefficient $\beta$ and hence $\Theta_{D}$. The $\Theta_{D}$ estimated from $\beta$ for few La-compounds are: $\Theta_{D}$ = 282~K for ${\rm La_2Co_3Ge_5}$, $\Theta_{D}$ = 304~K for ${\rm La_2Rh_3Ge_5}$, and $\Theta_{D}$ = 293~K for ${\rm La_2Pd_3Si_5}$.

\begin{figure} 
\centering
\includegraphics[width=8cm, keepaspectratio]{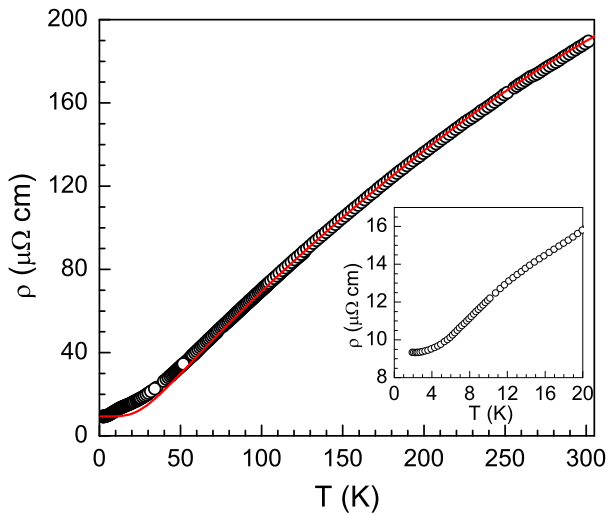}
\caption{\label{fig8} (Colour online) The temperature dependence of electrical resistivity of Pr$_{2}$Pt$_{3}$Si$_{5}$. The solid line represents the fit to Bloch-Gr\"{u}niesen-Mott relation discussed in text. The inset shows the expanded view of low temperature data.}
\end{figure}

The temperature dependence of the electrical resistivity of Pr$_{2}$Pt$_{3}$Si$_{5}$ is presented in Fig.~8. The electrical resistivity data exhibit a metallic behavior with a residual resistivity of $\rho_0$ = 9.3~$\mu \Omega$~cm at 2~K and a residual resistivity ratio RRR ($\rho_{300K}/ \rho_{2K}$) of $\sim$~20. The low value of residual resistivity and high value of residual resistivity ratio are measure of the good quality of our sample. Despite the presence of crystal field effect the electrical resistivity data of Pr$_{2}$Pt$_{3}$Si$_{5}$ could be well described within the framework of Boltzman transport theory using the Bloch-Gr\"{u}neisen-Mott relation,
\begin{equation}
\rho(T) = \rho_0 + \frac {4B}{\Theta_R} \left( \frac{T}{\Theta _R} \right)^5 \int_0^{\Theta _R/T} \frac{z^5 dz}{(e^z-1)(1-e^{-z})} - \alpha T^3 \nonumber
\end{equation}

\noindent where $\rho_0$ is the residual resistivity due to static defects in the crystal lattice, second term represents the contribution to resistivity due to electron-phonon scattering \cite{37} in which $\Theta_R$ is the Debye temperature obtained from resistivity data and $B$ is the electron-phonon coupling constant, and third term represents the contribution due to s-d interband scattering \cite{38,39}, $\alpha$ being the Mott coefficient. A fit to the electrical resistivity data with Bloch-Gr\"{u}neisen-Mott equation is represented by the solid line in Fig.~8. The best fit was obtained with parameters $\rho_0$ = 9.33~$\mu \Omega$~cm, $B$ = 28.78~m$\Omega$~cm~K, $\alpha$  = 1.02~$\times$~10$^{-6}$~$\mu \Omega$~cm/K$^3$, and Debye temperature $\Theta_R$ = 162~K. The slight departure of experimental resistivity data from the fit at low temperatures can be attributed to the presence of crystal electric field. The value of $\Theta_R$ = 162~K obtained from resistivity data is different from that obtained from the specific heat data, $\Theta_{D}$ = 252~K. The difference between the two values of Debye temperature arises due to the approximations made in Debye theory of lattice heat capacity that leads to a temperature dependent Debye temperature $\Theta_{D}$ which is higher at low temperatures, whereas the Bloch-Gr\"{u}neisen fit of electrical resistivity data gives an average value of Debye temperature $\Theta_R$ over an extended temperature range. Furthermore, the heat capacity was measured at constant pressure while Debye theory best describes the heat capacity at constant volume, this might also contribute to the difference between $\Theta_{D}$ and $\Theta_R$.

\section{Conclusions}

The magnetic and transport properties of Pr$_{2}$Pt$_{3}$Si$_{5}$ have been investigated. We found that Pr$_{2}$Pt$_{3}$Si$_{5}$ crystallizes in the U$_2$Co$_3$Si$_5$-type orthorhombic structure (space group {\it Ibam}). A van-Vleck paramagnetism is observed in the dc magnetic susceptibility data measured at a field of 0.05~T and above. The dc magnetic susceptibility data measured at low fields as well as the ac magnetic susceptibility data exhibit a sharp anomaly near 15~K whose origin is not understood.  However, the specific heat data do not show any anomaly near 15~K which suggests that there is no long range order in Pr$_{2}$Pt$_{3}$Si$_{5}$ down to 2~K. The Schottky-type anomaly observed in specific heat data reveals a CEF-split singlet ground state which we think is responsible for the absence of the long range magnetic order in Pr$_{2}$Pt$_{3}$Si$_{5}$. A single crystal study is desired to understand the origin of 15~K anomaly and hence the magnetic behavior of this compound.

\end{document}